\documentclass[batteries,article,accept,pdftex,moreauthors]{Definitions/mdpi}

\usepackage{amsmath}
\usepackage{amsfonts}
\usepackage{amssymb}

\firstpage{1} 
\pubvolume{1}
\issuenum{1}
\articlenumber{0}
\pubyear{2023}
\copyrightyear{2023}
\externaleditor{Academic Editor: {King Jet Tseng}}
\datereceived{01 December 2022} 
\daterevised{10 January 2023} 
\dateaccepted{13 January 2023} 
\datepublished{} 
\hreflink{https://doi.org/} % If needed use \linebreak
\Title{Data-Driven Thermal Anomaly Detection in Large Battery~Packs$^\dagger$}

\TitleCitation{Data-Driven Thermal Anomaly Detection in Large Battery Packs}

\Author{Kiran Bhaskar %MDPI: Please carefully check the accuracy of names and affiliations.
 $^{1}$\orcidA{}, Ajith Kumar $^{2}$, James Bunce $^{2}$, Jacob Pressman $^{2}$, Neil Burkell $^{2}$, Christopher D. Rahn $^{1,*}$\orcidB{}}

\AuthorNames{Kiran Bhaskar, Ajith Kumar, James Bunce, Jacob Pressman, Neil Burkell, Christopher D. Rahn}

\AuthorCitation{Bhaskar, %MDPI: Please check surname and given name carefully.
 K.; Kumar, A.; Bunce, J.; Pressman, J.; Burkell, N.; Rahn, C. D.}
\address{%
$^{1}$ \quad Department of Mechanical Engineering, The Pennsylvania State University, University Park, PA 16802, USA; kxb5761@psu.edu %MDPI: We added the email addresses here according to the submitting system. Please confirm.
\\
$^{2}$ \quad Wabtec Corporation, Pittsburgh, PA 15212, USA; ajith.kumar@wabtec.com (A.K.);\linebreak   james.bunce@wabtec.com (J.B.); jacob.pressman@wabtec.com (J.P.); neil.burkell@wabtec.com (N.B.)} 

\corres{Correspondence: cdr10@psu.edu

$^\dagger$ \quad This paper is an extended version of our paper published in the 2022 American Control Conference (ACC), Atlanta, GA, USA, 8--10 June 2022; pp. 5277--5281.
}

\abstract{The early detection and tracing of anomalous operations in battery packs are critical to improving performance and ensuring safety. This paper presents a data-driven approach for online anomaly detection in battery packs that uses real-time voltage and temperature data from multiple Li-ion battery cells. Mean-based residuals are generated for cell groups and evaluated using Principal Component Analysis. The evaluated residuals are then thresholded using a cumulative sum control chart to detect anomalies. The mild external short circuits associated with cell balancing are detected in the voltage signals and necessitate voltage retraining after balancing. Temperature residuals prove to be critical, enabling anomaly detection of module balancing events within $14$ min that are unobservable from the voltage residuals. Statistical testing of the proposed approach is performed on the experimental data from a battery electric locomotive injected with model-based anomalies. The proposed anomaly detection approach has a low false-positive rate and accurately detects and traces the synthetic voltage and temperature anomalies. The performance of the proposed approach compared with direct thresholding of mean-based residuals shows a $56\%$ faster detection time, $42\%$ fewer false negatives, and $60\%$ fewer missed anomalies while maintaining a comparable false-positive~rate.}


\begin{document}

%%%%%%%%%%%%%%%%%%%%%%%%%%%%%%%%%%%%%%%%%%
% The order of the section titles is different for some journals. Please refer to the "Instructions for Authors” on the journal homepage.

\section{Introduction}

Li-ion batteries (LiBs) are widely used in energy storage applications, such as power grids, electric vehicles, and electric locomotives, due to their high energy density, power density, long cycle life, and extended calendar life. Feng et al. \cite{b1}, however, list several recent accidents due to the failure of LiBs, often due to thermal runaway. Thermal runaway is often preceded by an internal short circuit caused by thermal, mechanical, and electrical abuse. Overcharge and over-discharge can also lead to thermal runaway \cite{b3}. Other critical anomalies in battery packs include balancing circuit failures and external short circuits (ESCs). Furthermore, sensor anomalies can lead to inaccurate control actions by the battery management system (BMS). Thus, it becomes critical to have an early and quick detection method followed by appropriate actions to avoid fault propagation, ensuring the safe and reliable operation of LiB packs.

The time-series data outputs of a battery system are non-stationary due to the time-varying current and environmental conditions. Anomalies may not be detected by directly thresholding the voltage and temperature measurements, especially at anomaly initiation when the voltage and temperature deviations are small. Therefore, the data are made stationary by estimating the voltage and temperature residuals as the difference between the measurements and the expected responses. Previous research focuses on cell-level anomaly detection using model-based residual estimation and thresholding \cite{b13,b8,b9,b10,b7,b11,b12}. State observers, such as extended Kalman filters (EKF) \cite{b35,b46}, adaptive EKF \cite{b12}, unscented Kalman filters (UKF) \cite{b33}, dual EKF \cite{b32}, and nonlinear observers \cite{b13}, have been used, along with parameter estimation techniques, such as recursive least squares \cite{b33,b7,b26,b46} and particle swarm optimization \cite{b11,b26}, to generate residuals. Anomaly detection is also performed by thresholding the model-based voltage, temperature, and state of charge (SoC) residuals against predetermined thresholds \cite{b32,b12,b14,b19}. \textls[-15]{However, generating model-based residuals is computationally expensive for battery packs, as it involves estimators for many cells. Computational complexity can be reduced through bar-delta filtering, cell mean models, and cell difference models to estimate the SoC of each cell \cite{b26,b25,b46}. Several works are reported in the literature that detect different types of battery-related \cite{b8,b9,b10,b7,b11} and sensor-related~\cite{b12,b35,b44,Sensor}} anomalies. \textls[-25]{However, most of the aforementioned approaches are applicable to only one type of fault~\cite{b7,b11,b12,b35}}. Some techniques work only if no two faults occur at the same time \cite{Sensor,Corr4,b19}. Some of the aforementioned approaches require parameter estimation by performing specific characteristic tests \cite{b10,b12,b44}.

Apart from model-based approaches, data-driven models, which utilize the cell-to-cell redundant voltage information in battery packs, are used for anomaly detection. Correlation-based methods detect and trace voltage anomalies using the correlation coefficient between cell voltages  \cite{Corr4,Corr1,Corr2}. However, these methods can be sensitive to measurement noise~\cite{Dey}. Entropy-based anomaly detection methods detect voltage anomalies by monitoring the entropy measure such as Shannon entropy \cite{Shannon1,Shannon2,Shannon3}. Sun et al. \cite{b20} detected and located short-circuit anomalies in battery packs by thresholding the modified Z-score of the relative entropy of individual cells with the pack median. Shannon entropy is also used for thermal runaway prognosis by detecting thermal faults \cite{ShannonTemp}. These methods have high computational costs and their performance is dependent on the choice of entropy measure and computation window, especially in the case of noisy data \cite{Dey}. Machine learning (ML)-based anomaly detection approaches that have been applied to other domains and LiBs \cite{b36} include classification, clustering, nearest-neighbor, statistical, information-theoretic, and spectral-based techniques \cite{b37}. ML techniques, such as neural networks \cite{b17}, the k-means clustering algorithm \cite{kmeans}, support vector machines \cite{b16}, and random forest classifiers \cite{b2,b18}, have also been applied to anomaly detection in battery systems. However, most of these techniques require large amounts of labeled battery-fault data for training.

Among the other data-driven approaches, Principal Component Analysis (PCA) is a promising unsupervised anomaly detection algorithm that has been extensively used in anomaly detection for multivariate systems \cite{b42,b40,b45}. Wang et al. \cite{b45}, for example, proposed sensor fault detection for a chiller system using PCA on the process variables. \mbox{Schmid et al. \cite{PCA}} proposed a PCA-based approach that detects voltage anomalies in a group of cells by applying PCA on voltage data processed using outlier robust sample studentization. In \cite{KPCA}, these researchers extended their method to include the kernel PCA-based method to detect internal short-circuit (ISC) faults using voltage signals but it is computationally expensive. The approaches in \cite{PCA,KPCA} detected anomalies with a single anomalous voltage but their applicability in the case of multiple anomalous signals was not studied. The effect of cell balancing on detection performance was also not studied. Furthermore, the literature lacks an effective anomaly detection approach that can also detect thermal anomalies, even in the case of multiple anomalous signals.

This paper extends and improves the work in \cite{PCA} to present an anomaly detection scheme that combines PCA and the cumulative sum (CUSUM) control chart to detect and locate both voltage and temperature anomalies in groups of Li-ion cells in real time. Addressing the aforementioned research gaps, the proposed approach detects voltage and temperature anomalies, even in the case of multiple simultaneous anomalous signals. The voltage and temperature residuals are the difference between the measured cell signals and the mean signals of the cell group. Unlike model-based approaches, median-based and mean-based residuals (MBRs) reduce the effect of aging, as all the cells in the cell group experience similar loading and environmental conditions during their life. In the proposed approach, the MBRs are processed using PCA to capture cell-to-cell information including the inconsistencies and thresholded using the CUSUM control chart to detect anomalies, which reduces the false positives and improves the detection rate. Experimental validation of the proposed approach is performed on external short-circuit data from a battery electric locomotive. We compare the proposed approach using PCA-processed MBRs (PCA method) with the direct thresholding of voltage and temperature MBRs (direct method)~\cite{KB}. To further evaluate performance, statistical testing of the proposed approach is performed using model-based synthetic anomalies injected into nominal experimental data. The detection time, recovery time, false-negative rate, missed anomaly rate, and false-positive rate statistics are compared for the two methods. Finally, the effect of balancing on the performance of the anomaly detection approach is also studied. 
%%%%%%%%%%%%%%%%%%%%%%%%%%%%%%%%%%%%%%%%%%

\section{Anomaly Detection Algorithms}

Mean-based residual generation is proposed for both voltage and temperature. We assume that the cells within a group behave similarly under nominal conditions. These cell groups could be battery strings consisting of series/parallel connected cells that are spatially, thermally, chemically, and electrically similar. The MBRs of voltage and temperature in a cell group with $n$ cells are calculated by
\begin{equation}
x_i(t)=X_i(t)-\mu_{X}(t)
\label{eq1}
\end{equation}
where ${X}_{i}$ is the voltage/temperature of the ${i}^{th}$ cell and the mean $\displaystyle \mu_{X}(t)=\frac{1}{n}\sum_{i=1}^{n} X_i(t)$. 
\\

\begin{figure}[H]
\centerline{\includegraphics[width=\textwidth]{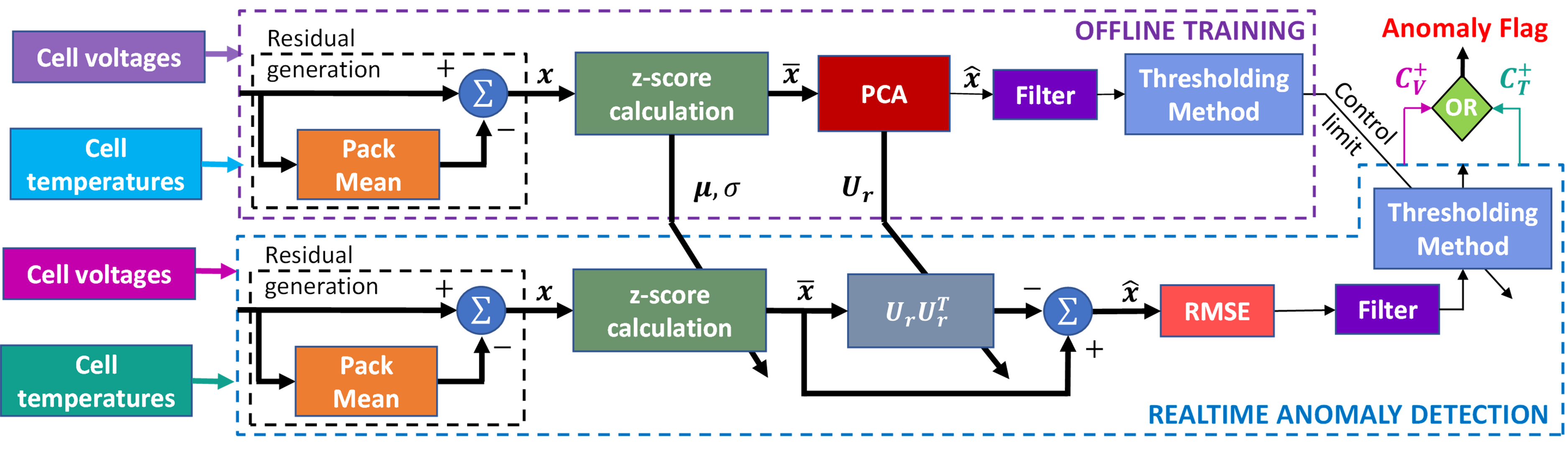}}
\caption{Block diagram of proposed PCA-based anomaly detection algorithm (PCA Method).}
\label{PCA_schematics}
\end{figure}

The first and simplest anomaly detection scheme (direct method) directly compares each residual signal with a predetermined threshold \cite{KB}. For earlier anomaly detection, the PCA method captures cell-to-cell heterogeneity using PCA. Figure \ref{PCA_schematics} shows a block diagram for anomaly detection using the PCA method. Both methods rely on parameters derived from $k$ samples of the residuals of nominal anomaly-free data. This training data provide the mean of the voltage/temperature residuals of each cell ($\mu_{Vr,i}$/$\mu_{Tr,i}$) and the standard deviation of the voltage/temperature residuals for all cells ($\sigma_{Vr}$/$\sigma_{Tr}$), which are used to calculate the Z-score. PCA is applied on the Z-score because the application of PCA directly on residuals would point the first principal component toward the mean of the data instead of the direction of the highest variance of the residuals.

%\\

The training data are placed in the matrix $\textbf{X}\in \mathbb{R}^{n\times{k}}$ and decomposed via singular value decomposition to $\textbf{X=USV}^T$, where $\textbf{U}\in\mathbb{R}^{n\times{n}}$ is the left singular matrix, $\textbf{S}\in\mathbb{R}^{n\times{n}}$ is the singular value matrix, and $\textbf{V}\in\mathbb{R}^{k\times{n}}$ is the right singular matrix. The number of principal components, $p$, is selected to provide a cumulative variance of $90\%$ \cite{b40}. The truncated left singular matrix $\textbf{U}_r$ is the first $p$ columns of $\textbf{U}$.

In real time, $\textbf{x}(t)$ is measured, mean shifted, and normalized by the training data $\mu_{Xr,i}$ and $\sigma_{Xr}$ to estimate the Z-score as
\begin{equation}
    \overline{\textbf{x}}(t)=\frac{x_i(t)-\mu_{Xr,i}}{\sigma_{Xr}} 
\end{equation}
\noindent
The matrix multiplication $\textbf{U}_r^T \overline{\textbf{x}}(t)$ is the projection into the lower dimensional space (i.e., principal subspace), where only nominal data points exist, and $\hat{\textbf{x}}(t)=\textbf{U}_r \overline{\textbf{x}}(t)$ gives the projection back to the original dimension. The reconstruction of $\overline{\textbf{x}}(t)$ \cite{b22} is
\begin{equation}
\hat{\textbf{x}}(t)=\textbf{U}_r\textbf{U}_r^T \overline{\textbf{x}}(t).
\end{equation}
\noindent 
Because the principal subspace spanned by $\textbf{U}_r$ is anomaly-free, the reconstruction $\hat{\textbf{x}}(t)$ is the expected value in the case of nominal operation. Therefore, an anomaly can be detected by monitoring the difference between $\overline{\textbf{x}}(t)$ and $\hat{\textbf{x}}(t)$.

Statistical process control (SPC) charts have been widely used in residual-based anomaly detection for stationary processes. Shewhart, CUSUM, and exponentially weighted moving average (EWMA) control charts are commonly used in univariate SPC \cite{b42,b21}. Among these, the CUSUM control chart is one of the most effective in detecting small deviations in monitored signals \cite{b42,b21}. CUSUM control charts have been used in model-based anomaly detection for battery systems \cite{b12,b35,KB}.

The PCA method reduces the normalized residual vector to a scalar using the RMSE and uses CUSUM statistics \cite{b21} to threshold the filtered RMSE. A simple first-order low-pass filter with a cutoff frequency manually tuned to $4.9$ mHz is used to filter the RMSE for robust detection by filtering out high-frequency variations. However, the direct method directly thresholds the absolute values of the filtered voltage and temperature residuals using CUSUM statistics, where the filter has a cutoff frequency of $8.4\, mHz$. The positive deviation CUSUM is 
$C^{+[t]}=max (0,C^{+[t-1]}+(y[t]-{\mu_c})-K)$, and the negative deviation CUSUM is $C^{-[t]}=max (0,C^{-[t-1]}-(y[t]-{\mu_c})-K)$, with $C^+[0]=C^-[0]=0$ and $K$ is chosen to be $4\sigma_c$ for lower false positives, where $\mu_{c}$ and $\sigma_{c}$ are the mean and standard deviation of the thresholding variable, $y[t]$, for the anomaly-free training data. Both $C^+$ and $C^-$ are compared against the $5\sigma_c$ control limits for the direct method \cite{KB}. In the PCA method, it is sufficient to compare $C^+$ against the $5\sigma_c$ control limit because PCA always produces a positive deviation in the RMSE. Voltage and temperature anomalies are detected independently from $C^+_V$ and $C^+_T$, respectively.

If an anomaly is detected, the anomalous cell can be identified as the cell with the maximum absolute error in the reconstructed residuals. The first and first two principal components are used to reconstruct the voltage and temperature residuals, respectively, for good tracing performance \cite{b22}. 

\section{Synthetic Anomalous Data}

LiBs have been commonly modeled using electrochemical and equivalent circuit models (ECM). The former are more accurate and explain the electrochemical processes that occur inside a battery but are computationally expensive \cite{RahnText}. The latter are computationally efficient and can provide sufficient accuracy to be widely used in real-time applications~\cite{b23}. Thevenin's equivalent circuit models have been widely used to model LiBs \cite{b13,b33,b32}, sometimes including a short-circuit resistance to model LiB cells under internal short circuits~\cite{b8,b29,b30}. Higher-order dynamic thermal models are available in the literature~\cite{b9,b10} but a lumped thermal model is often sufficiently accurate \cite{b13}.

\begin{figure}[H]
\centerline{\includegraphics[width=\linewidth]{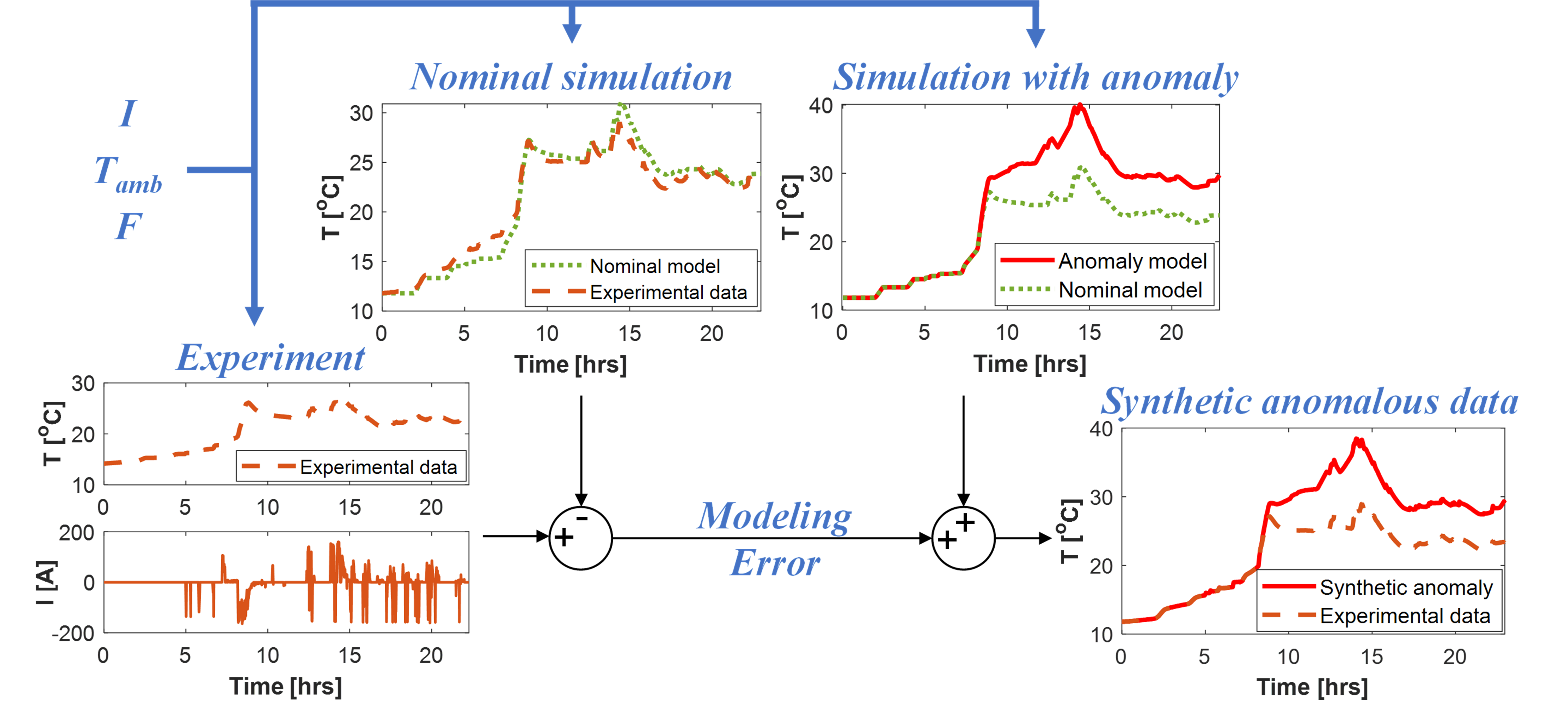}}
\centering
\caption{Flow chart for synthetic data generation for an air$-$flow anomaly.}
\label{SynAno}
\end{figure}

One of the main challenges in testing anomaly detection algorithms is the lack of experimental anomalous data. We adopted a hybrid experimental model approach rather than relying exclusively on model-based data. Anomalies are injected into the voltage and/or temperature of the anomalous cell. Two sensor anomalies, loose voltage and temperature sense leads, are injected by adding bias terms with noise into the experimental data. Figure~\ref{SynAno} shows the schematics of the anomaly injection approach to create synthetic anomalous data for internal short circuits, air flow anomalies, and voltage dropouts. Thevenin’s equivalent circuit model with short circuit resistance and a first-order lumped thermal model are used to generate anomalous voltage and temperature data, respectively. In the electrical model, the state propagation for the SoC $(z)$ and diffusion voltage $(V_c)$, and output equations~\cite{KB}~are:

\begin{equation}
\begin{gathered}
    I_b(t)=I_{sc}(t)+I(t), \\[2 pt]
    \dot{z}(t)=-\frac{I_b(t)}{36Q}, \\[2 pt]
    \dot{V}_{c}(t)=-\frac{1}{R_{1} C_{1}} V_{c}(t)+\frac{I_b(t)}{C_{1}},  \\[3 pt]
    V(t)=\frac{{OCV}(z(t))-V_{c}(t)-I(t)R_{0}}{R_0+R_{sc}} R_{sc},
\end{gathered}
\label{eq3}
\end{equation}

\noindent
where $R_0$ is the Ohmic resistance, $R_1$ is the polarization resistance, $C_1$ is the polarization capacitance, $Q$ is the capacity, $R_{sc}$ is the short-circuit resistance, and $OCV$ is the open circuit voltage as a linear function of $z$. The thermal behavior can be modeled using the dynamic model \cite{KB}, 

\begin{equation}
  \begin{aligned}
     \dot{T}(t)=a\left(I(t)^{2} R_{0} + I_{sc}(t)^2 R_{sc} + \frac{V_{c}(t)^{2}}{R_{1}}\right) +b\left(T(t)-T_{amb}(t)\right)F(t),
    \label{eq4}    
  \end{aligned}
\end{equation}
\noindent{where $a$ is the thermal dissipation coefficient and $b$ is the thermal inertia coefficient. ${T}_{amb}$ is the ambient temperature and $F$ is the fan status (0 for off and 1 for on). This thermal model incorporates heat generation due to the ISC modeled as Joule's heating \cite{ThermalISC}.}

As the proposed approach is a purely data-driven approach, it essentially detects the cell-to-cell voltage and temperature deviations from the nominal setting and does not depend on the cause of the deviation. Therefore, synthetic anomalous data are generated to test the effectiveness of the proposed approach for different fault signatures. For example, even though a real air$-$flow anomaly could lead to multiple anomalous temperatures, we inject an anomaly only in one cell temperature to test the detection performance because the higher the number of anomalous signals, the easier they can be detected by PCA. Hence, statistical testing using multiple cell anomalies may not be ideal for evaluating detection performance and anomalous cell identification accuracy.

Six performance indices are used to evaluate anomaly detection performance: the detection time ($DT$), recovery time ($RT$), false-negative rate ($FNR$), false-positive rate ($FPR$), missed anomaly rate ($MAR$), and true tracing rate ($TTR$). The $DT$ is the time between the start and detection of the anomaly. The $RT$ is the elapsed time between the end and flag reset of the anomaly. The FNR is the percentage of false negatives between the first detection and the end of the anomaly. The $FPR$ is the percentage of the time that the anomalies are flagged in the nominal data. The $MAR$ is the ratio of missed detections to total anomalies. The true tracing rate ($TTR$) is the percentage of time the anomalous cell is located accurately when an anomaly is detected. Thus, an ideal approach will have a low $DT$, low $RT$, low $FPR$, low $FNR$, low $MAR$, and a high $TTR$.

\section{Results and Discussion}
\subsection{Battery System and Data}
 
This study used experimental current, voltage, temperature, and fan-status data from a Wabtec FLXDrive battery electric locomotive battery pack consisting of 825 Li-ion NMC cells with a $37$ Ah capacity in a 275S-3P arrangement. The 3P cells were considered a single equivalent cell with the same voltage, three times the capacity, and each cell receiving $1/3$ of the current. Twenty-five cell groups were formed, each with 11 similar cells. The voltage ($V$) and surface temperature ($T$) were measured for each cell and the current ($I$) was measured for the entire battery pack (VTI data) during nominal locomotive operations. The current was positive during discharging and negative during charging. The battery pack was air-cooled and a fan blew air into the pack to enhance the convective heat transfer. The ambient temperature ($T_{amb}$) and fan status ($F$) were also measured for each sub-group. All the measurements were sampled at 1 Hz. During cell balancing, a passive circuit discharged the cells to the lowest SoC within the series string through a shunt resistance of 100 $\Omega$.

\subsection{Validation of Synthetic Anomalous Data}
\begin{figure}[H]
\centerline{\includegraphics[width=\textwidth]{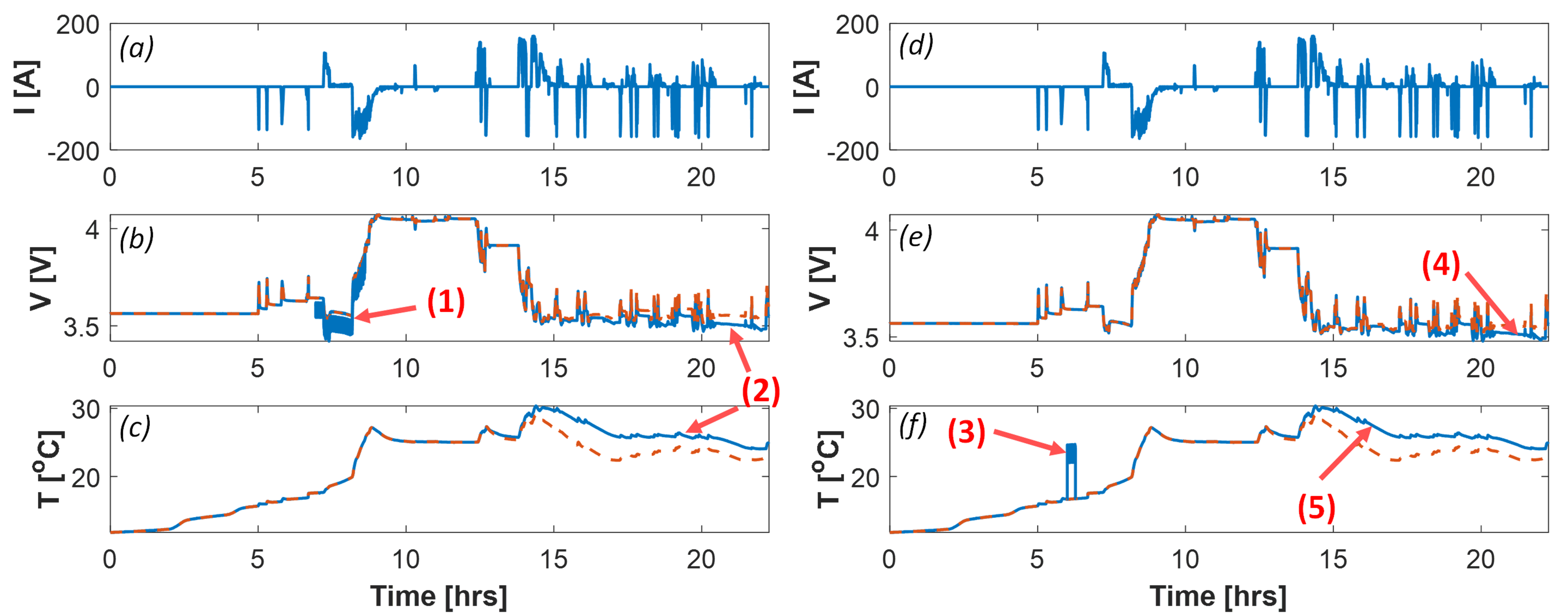}}
\caption{Examples of nominal (dashed) and synthetic anomalous (solid) voltage (\textbf{b},\textbf{e}) and temperature (\textbf{c},\textbf{f}) data associated with (1) loose voltage sense lead, (2) ISC, (3) loose temperature sense lead, (4) voltage dropout, and (5) air$-$flow anomaly, with the input current (\textbf{a},\textbf{d}).}
\label{ex}
\end{figure}

\begin{figure}[H]
{\includegraphics[width=0.7\linewidth]{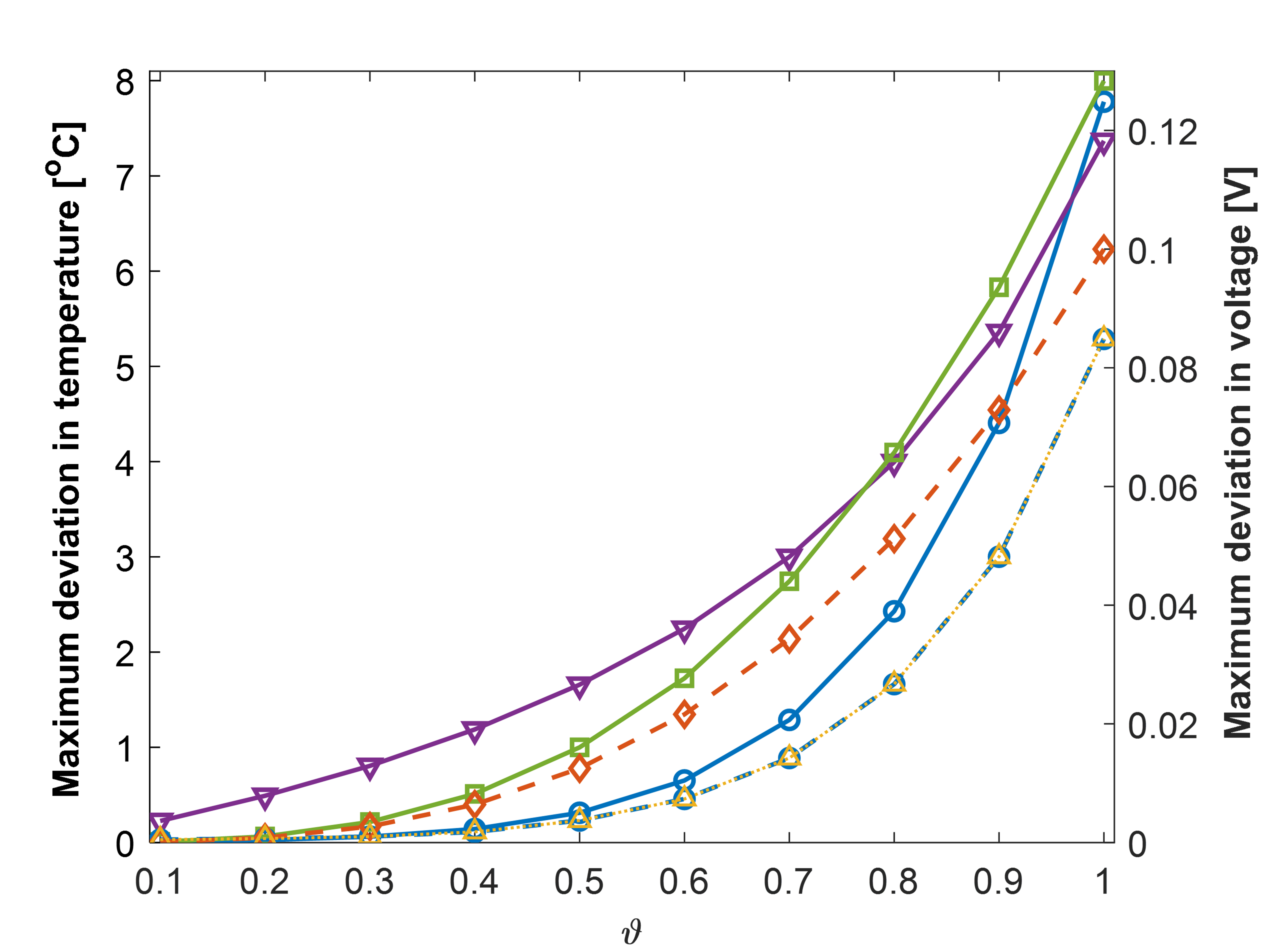}}
\caption{Variation in maximum deviation of temperature and voltage with anomaly magnitude; solid lines represent temperature deviations (left axis) and dashed/dotted lines represent voltage deviations: $\bigcirc$ ISC, $\bigtriangledown$ air flow, $\Box$ loose temperature sense lead, $\Diamond$ loose voltage sense lead, and $\bigtriangleup$ voltage dropout.}
\label{MaxDevVSMag}
\end{figure}

The model parameters were the batch least-square estimates, $R_0$, $R_1$, $C_1$, $Q$, $a$, and $b$. The anomaly model parameters were the anomaly type, anomaly magnitude ($\vartheta$), start time ($t_a$), and anomaly duration ($\Delta t_a$). The anomaly magnitude ranged from $0$ to $1$ from no anomaly to the most severe anomaly considered, respectively. The ISC was modeled with $R_{sc}=\exp{(9(1-0.6\vartheta)^2)}-1$ (See Figure \ref{ex}b). The voltage dropout anomaly was modeled with $R_{sc}=\exp{(9(1-0.6\vartheta)^2)}-1$ (See Figure \ref{ex}d). The air$-$flow anomaly was modeled with ${b}=(1-\vartheta)\hat{b}$, where $\hat{b}$ is the nominal value of the thermal inertia coefficient (See Figure \ref{ex}e). Unlike the other anomalies, loose voltage and temperature sense leads were injected for a fixed duration, as shown in Figures \ref{ex}a and \ref{ex}c, respectively. Figure \ref{MaxDevVSMag} shows that the maximum voltage and temperature deviations as functions of the anomaly magnitude for all five anomalies were similar. 

\subsubsection{Air$-$flow Anomaly Affecting a Single Cell's Temperature}
\label{airflowsingle}
\begin{figure}[H]
{\includegraphics[width=0.8\linewidth]{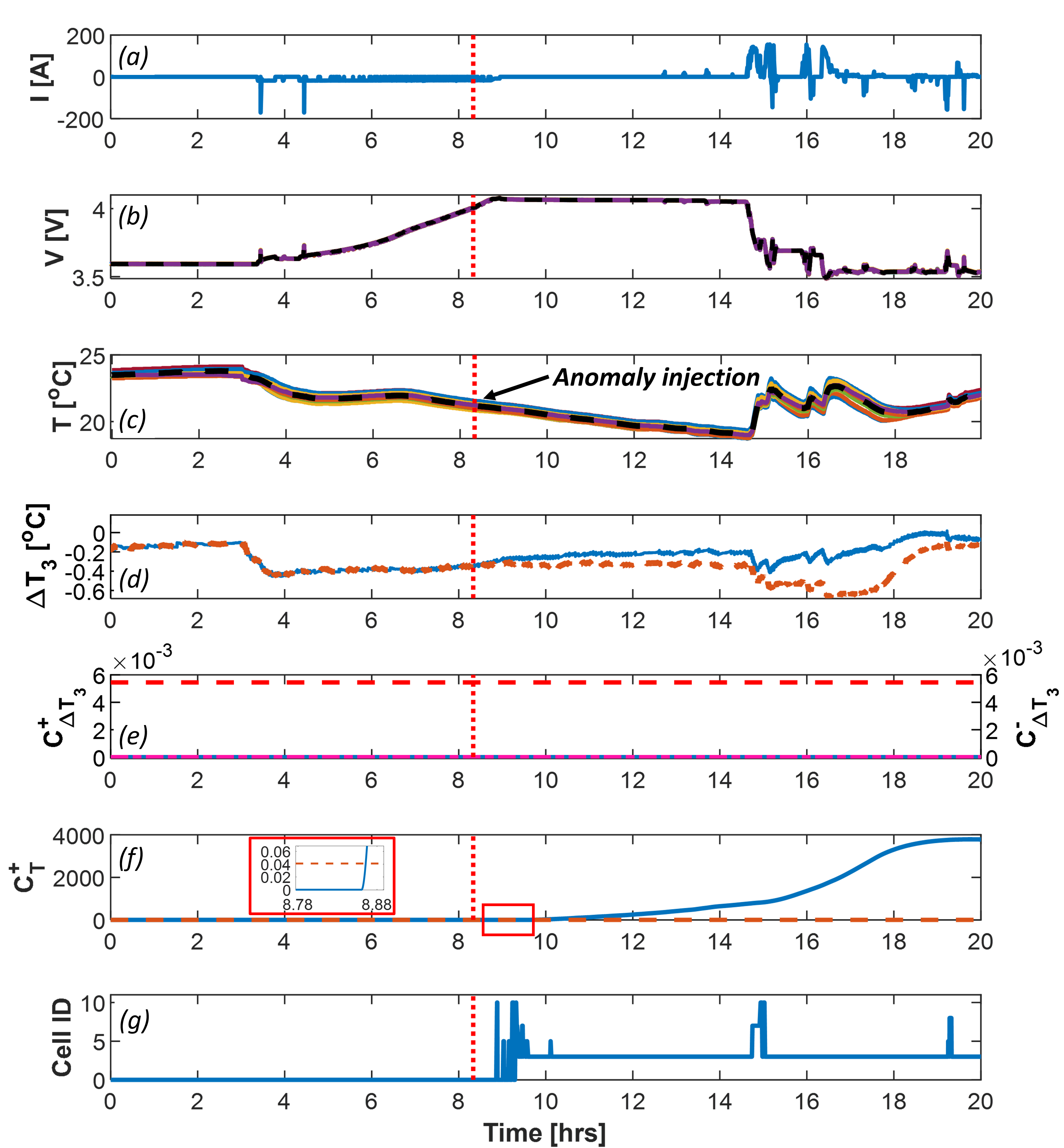}}
\caption{Anomaly detection using PCA (PCA method) and baseline (direct method) approaches for air$-$flow anomaly ($\vartheta=0.2$), with anomaly initiation (dotted) at 8.33 h: (\textbf{a}) Input current profile; (\textbf{b})~Voltage of 11 cells and mean voltage (dashed); (\textbf{c})  Temperature of 11 cells and mean temperature (dashed); (\textbf{d}) Nominal (dashed) and anomalous (solid) temperature MBR of cell $3$;  (\textbf{e}) Direct method temperature $C^+$ and $C^-$ of cell $3$ and threshold (dashed); (\textbf{f}) Temperature anomaly score from PCA method with threshold (dashed); and (\textbf{g}) PCA method tracing of anomalous cell.}
\label{AirFlow}
\end{figure}

\begin{figure}[H]
{\includegraphics[width=0.6\linewidth]{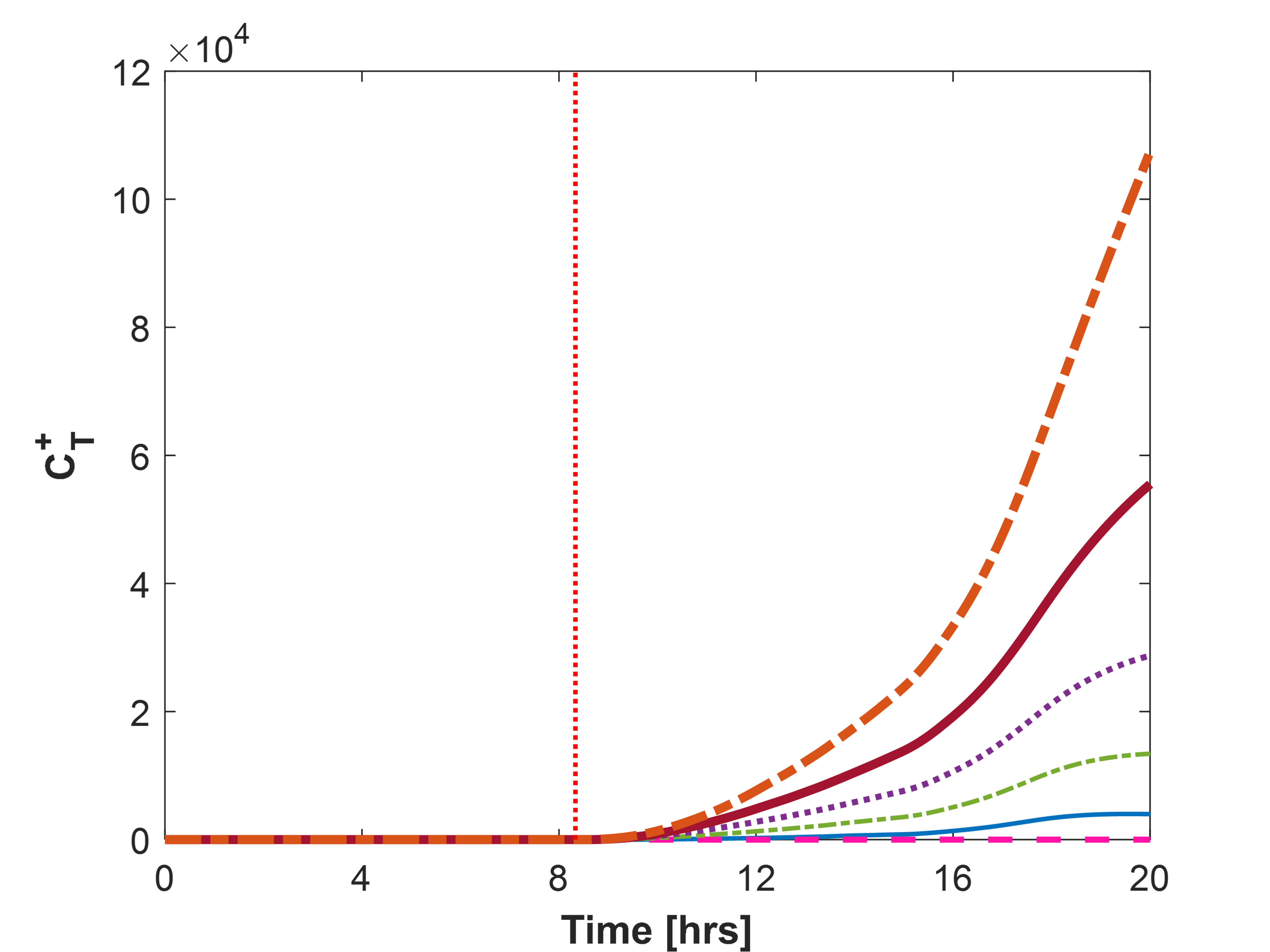}}
\caption{Variation in temperature anomaly score with anomaly magnitude for air$-$flow anomaly with anomaly initiation (dotted) at 8.33 h: Threshold (dashed), $\vartheta$ = 0.2 (solid), $\vartheta$ = 0.4 (dash-dotted), \mbox{$\vartheta$ = 0.6} (dotted), $\vartheta$ = 0.8 thick(solid), $\vartheta$ = 1 (thick dash-dotted).}
\label{Airflow_diffmag}
\end{figure}

Both anomaly detection algorithms were trained with 24 h of data and tested on a different 24 h of data. Figure \ref{AirFlow} shows an example of anomaly detection on a cell group with a mild ($\vartheta=0.2$) air$-$flow anomaly injected into cell $3$ at $t_a=8.33$ h,  with a deviation of $\Delta t_a=11.67$ h. The parameters obtained from the training process in the PCA method are reported in Table \ref{Training}. The nominal current is shown in Figure \ref{AirFlow}a. The voltage and temperature of the cells were tightly clustered, as shown in Figures \ref{AirFlow}b and \ref{AirFlow}c, respectively, before the anomaly was injected. The anomalous MBR of the cell $3$ temperature was smaller than its nominal MBR, as shown in Figure \ref{AirFlow}d. Figure \ref{AirFlow}e shows that $C^+_{\Delta T_3}$ and $C^-_{\Delta T_3}$ did not cross their thresholds. Thus, the direct method failed to detect this mild anomaly. Figure \ref{AirFlow}f shows the anomaly being detected using the PCA method, as the temperature anomaly score $C^+_T$ crossed the threshold around $33$ min after the anomaly injection. $C^+_T$ increased gradually, indicating a persistent anomaly. Figure \ref{AirFlow}g shows the corresponding anomalous cell being located with $90.9\%$ accuracy. Figure \ref{Airflow_diffmag} shows the anomaly score increasing nonlinearly with the anomaly magnitude. Thus, larger anomalies were substantially easier to detect.

\begin{table}[H]
\centering
\caption{{Outputs from training process in PCA method.}}

\begin{tabular}
{m{2.20cm}m{2.20cm}m{2.20cm}m{2.20cm}}
  \toprule
  \multicolumn{2}{c} {\textbf{Voltage PCA}} & \multicolumn{2}{c} {\textbf{Temperature PCA}}\\
   \midrule
  $\textit{p}_V$ & \quad\quad 7 & $\quad \textit{p}_T$ & \quad\quad 4\\

  $\sigma_{Vr} \,$ [V] & \quad 0.0018  &  $\quad \sigma_{Tr} \,$ [\textdegree C] & \quad 0.3205 \\

  $\mu_{c,V}$ & \quad 0.0423 & \quad $\mu_{c,T}$ & \quad 0.0353 \\ 
  
  $\sigma_{c,V}$ & \quad 0.0366 & $\quad \sigma_{c,T}$  & \quad 0.0082 \\
  
  $\textit{V}_{threshold}$ & \quad 0.1830 & $\quad \textit{T}_{threshold}$ & \quad 0.0410\\
  \bottomrule
\end{tabular}
\label{Training}

\end{table}

\vspace{-12pt}

\subsubsection{Air$-$flow Anomaly Affecting Two Cells' Temperatures}
\begin{figure}[H]
{\includegraphics[width=0.7\linewidth]{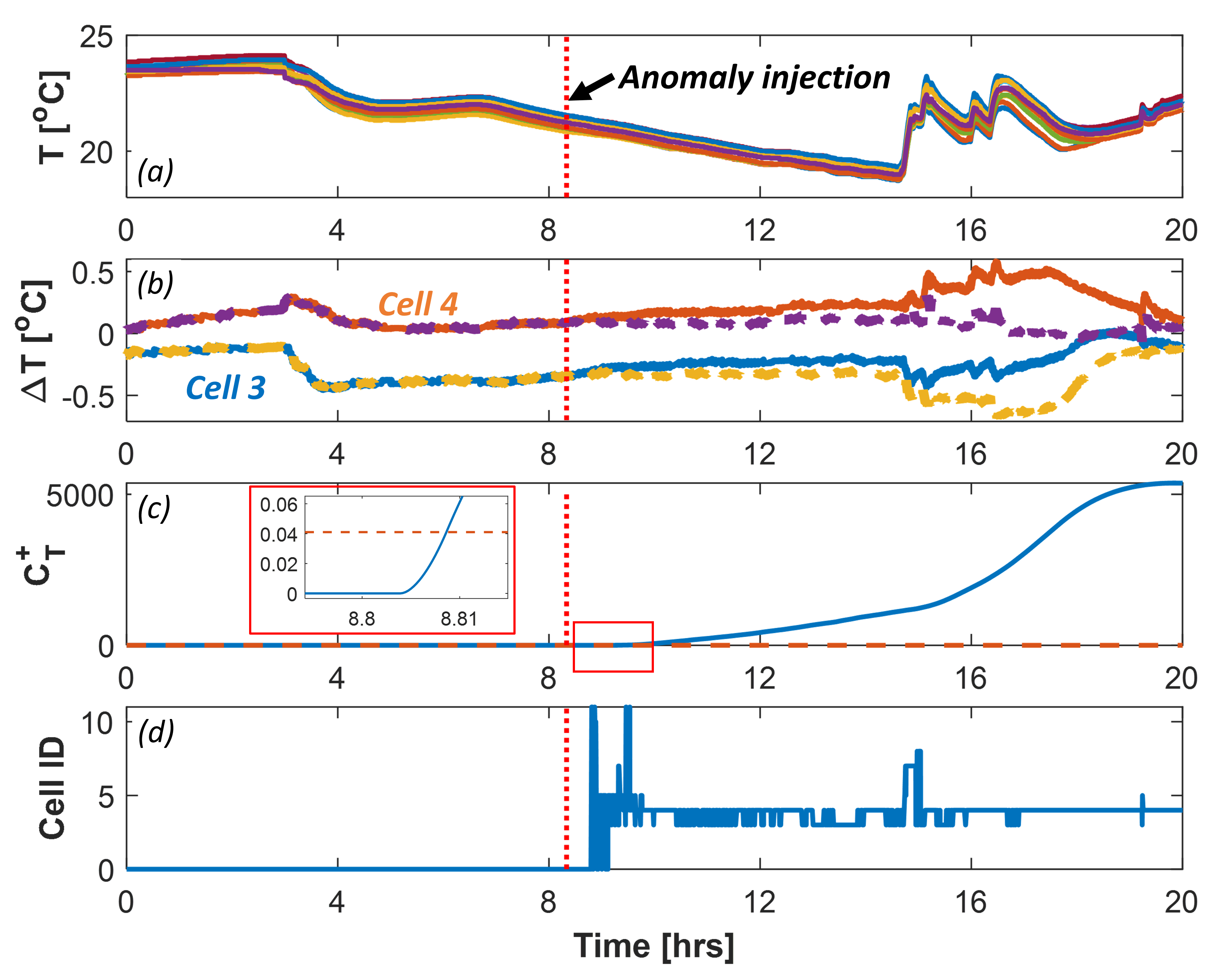}}
\caption{Detection of air$-$flow anomaly ($\vartheta=0.2$), leading to anomalous temperatures in cells $3$ and $4$, with anomaly initiation (dotted) at 8.33 hours, using PCA method: (\textbf{a}) Temperature of 11 cells and mean temperature (dashed); (\textbf{b}) Nominal (dashed) and anomalous (solid) temperature MBR of cells $3$ and $4$; (\textbf{c}) Temperature anomaly score from PCA method with threshold (dashed); and (\textbf{d}) PCA method tracing of anomalous cell.}
\label{2Airflow}
\end{figure}

To evaluate the ability of the proposed algorithm in the case of simultaneous multiple anomalous signals, the PCA method was tested on synthetic anomalous data, with a mild air$-$flow anomaly ($\vartheta=0.2$) initiated at $t_a=8.33$ h, leading to anomalous temperatures in cells $3$ and $4$ for $11.67$ h, as shown in Figure \ref{2Airflow}. The nominal experimental data are the same as in Section~\ref{airflowsingle}. Even though cells $3$ and $4$ had temperature anomalies, all the cell temperatures were tightly packed, as shown in Figure \ref{2Airflow}a, and thus undetectable by the direct method. The deviations from the nominal behavior were small and are visualized by comparing the MBRs in the anomalous and nominal cases in Figure \ref{2Airflow}b. The PCA method detected the anomaly within $29$ min (see Figure \ref{2Airflow}c) and traced the anomalous cell accurately as either cell $3$ or $4$ for more than $92.7\%$ of the time. It should be noted that the detection time, in this case, was $4$ min lower than that reported in Section~\ref{airflowsingle}, where only cell $3$ was anomalous. As evident by the detection times in both cases, the occurrence of multiple anomalous signals made them easier for the PCA method to detect because the cumulative change in the cell-to-cell relationship was amplified. Thus, the PCA method can detect anomalies in the case of multiple anomalous cells in the early stage and accurately trace the cell that has the most anomalous deviation. However, this approach failed to detect anomalies where all the cells within the cell group showed identical anomaly signatures, but this situation is highly unlikely in a real battery pack. For example, the voltage signals in the case of a module ESC tracked each other, but the temperature anomaly signatures were non-identical.   

\subsection{Experimental ESC testing}

\subsubsection{Testing on Single-Cell ESC}
\begin{figure}[H]
{\includegraphics[width=0.7\linewidth]{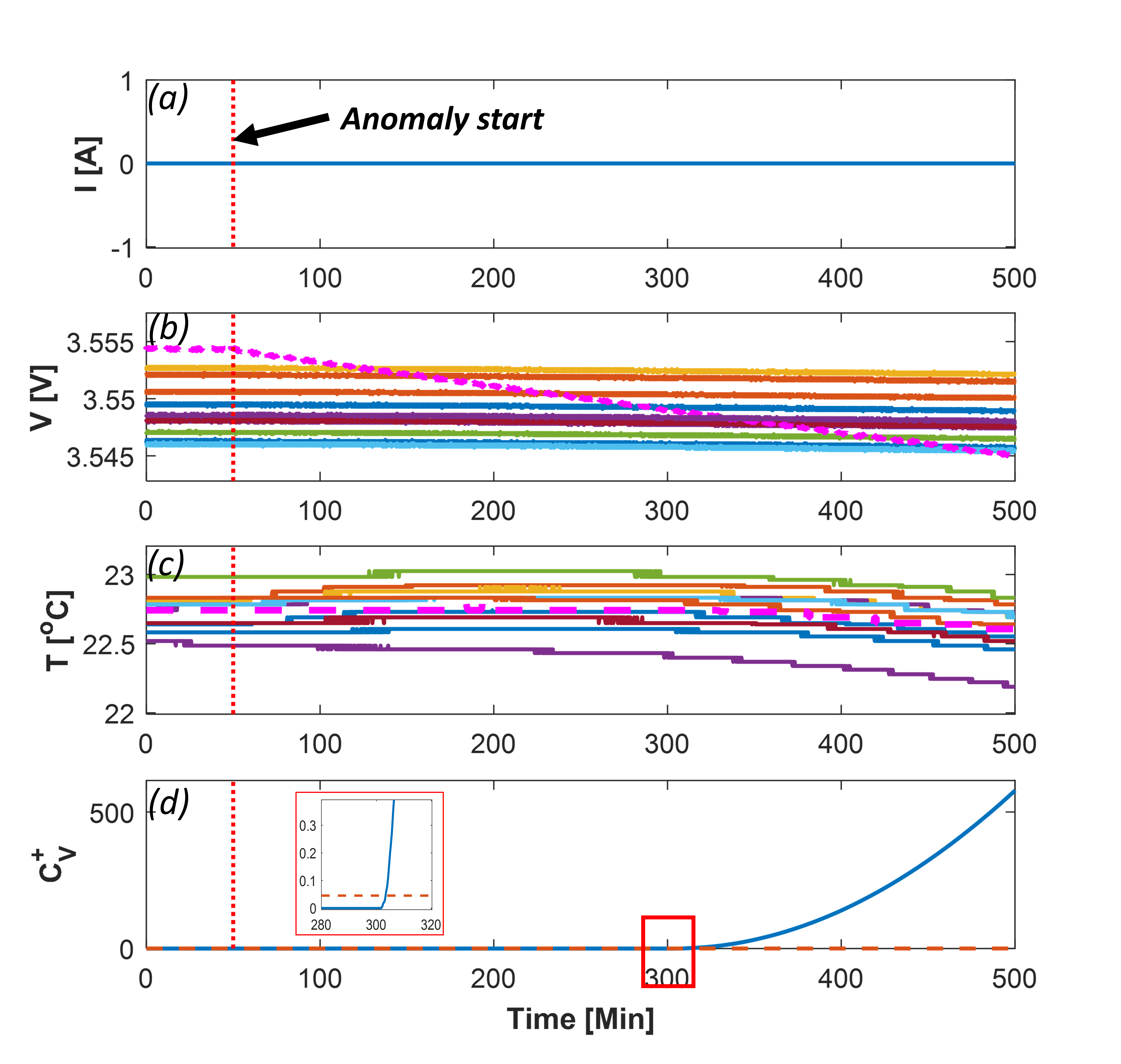}}
\caption{Detection using PCA method on experimental data with ESC in cell 10 initiated at 50 min (vertical, dotted): (\textbf{a}) Pack current; (\textbf{b}) Voltages of 11 cells (cell 10, dashed); (\textbf{c})  Temperatures of 11 cells (cell 10, dashed); (\textbf{d}) $C^+_V$  with its threshold (dashed)}
\label{SinglecellESC}
\end{figure}

Cell balancing in the battery electric locomotive involves connecting a 100 $\Omega$ shunt resistor across the cells' terminals. This is a mild ($\vartheta=0.47$) ESC or micro-short circuit~\cite{b26}. One single cell-balancing event and 13 module balancing events were used to evaluate the PCA method for ESC detection. Figure \ref{SinglecellESC} shows the single-cell ESC fault initiated at 50~min, causing cell $10$'s voltage (dashed line) to drop while the current was zero. The PCA method on the voltage data detected the ESC within $255$ min. The algorithm did not detect anomalies in the temperature data. The anomalous cell $10$ was accurately traced. Even though the example in Figure \ref{SinglecellESC} shows the application of the PCA method to a zero-current operation, this method did not use the current signals and detected the ESC, even when the current was non-zero because the residual of the shorted cell voltage behaved differently compared to the nominal cell-to-cell relationship.

\subsubsection{Statistical Testing on Module ESC}
\begin{figure}[H]
{\includegraphics[width=0.7\linewidth]{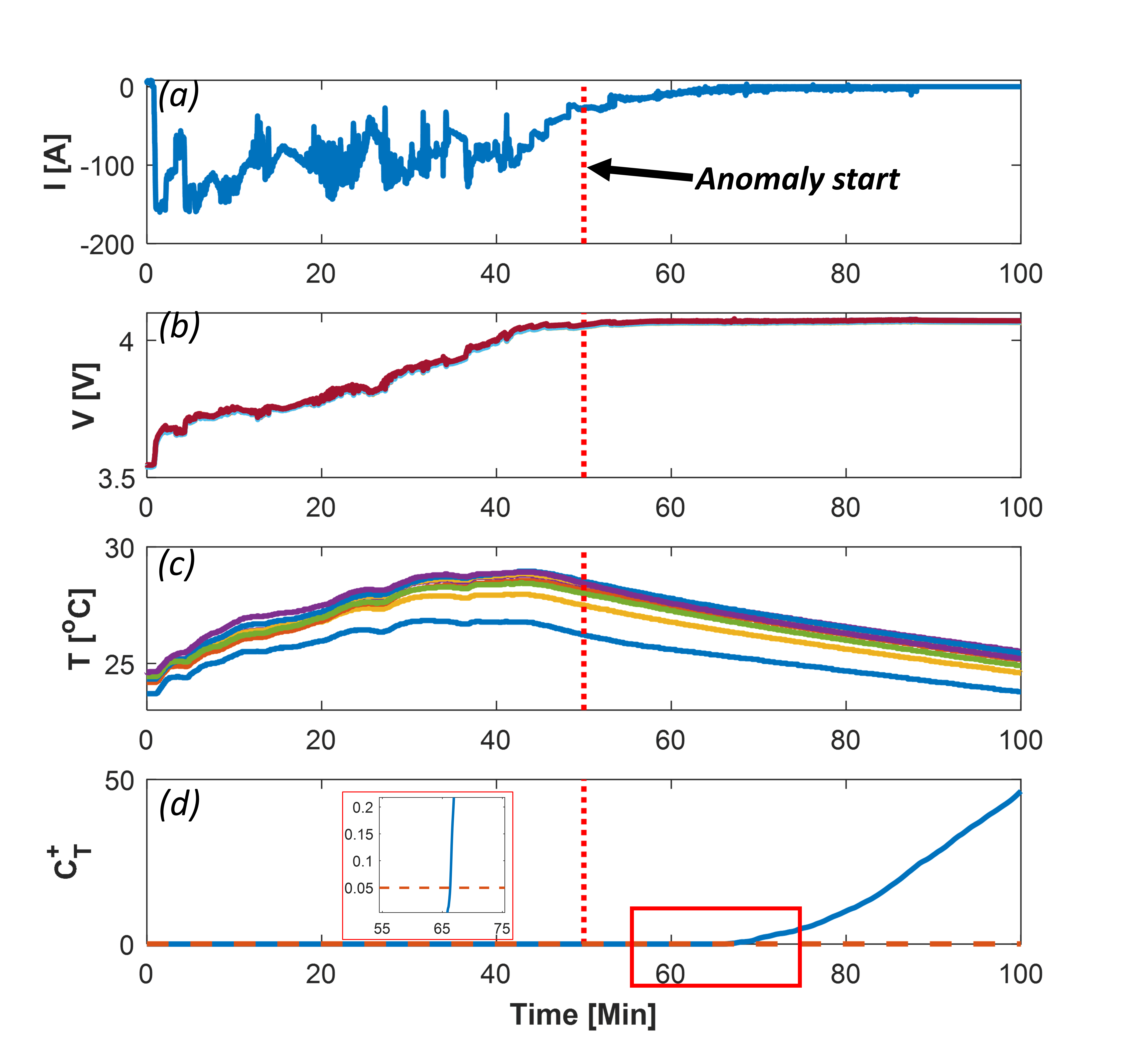}}
\caption{Detection using PCA method on experimental data with ESC in all cells initiated at 50 min (vertical dotted): (\textbf{a}) Pack current; (\textbf{b}) Voltage of 11 cells; (\textbf{c})  Temperature of 11 cells; (\textbf{d}) $C^+_T$  with its threshold (dashed).}
\label{ExpVal}
\end{figure}

During module balancing, all 11 cells experienced ESCs. The pack current, cell voltage, and cell temperature are shown in Figures \ref{ExpVal}a, \ref{ExpVal}b, and \ref{ExpVal}c, respectively. Figure \ref{ExpVal}d shows that the temperature PCA detected the anomaly within $16.3$ min. The voltage PCA, however, did not detect the anomaly because all the cells were balancing, as discussed earlier. In this example, even though the temperature variations were unnoticeable, as seen in Figure \ref{ExpVal}c, due to the cell-to-cell variation in the thermal dynamics, multiple anomalous temperatures were present. Statistical testing on 13 different module balancing events showed that the temperature PCA detected the fault within $13.5$ min, on average, with an $FNR$ of $2.3\%$. The voltage PCA, however, was ineffective with a $99\%$ $FNR$. This experimentally validates the ability of the PCA method to detect anomalies in the case of multiple anomalous signals.      

\subsection{Statistical Testing Using Synthetic Anomalous Data}
\begin{figure}[H]
{\includegraphics[width=0.8\linewidth]{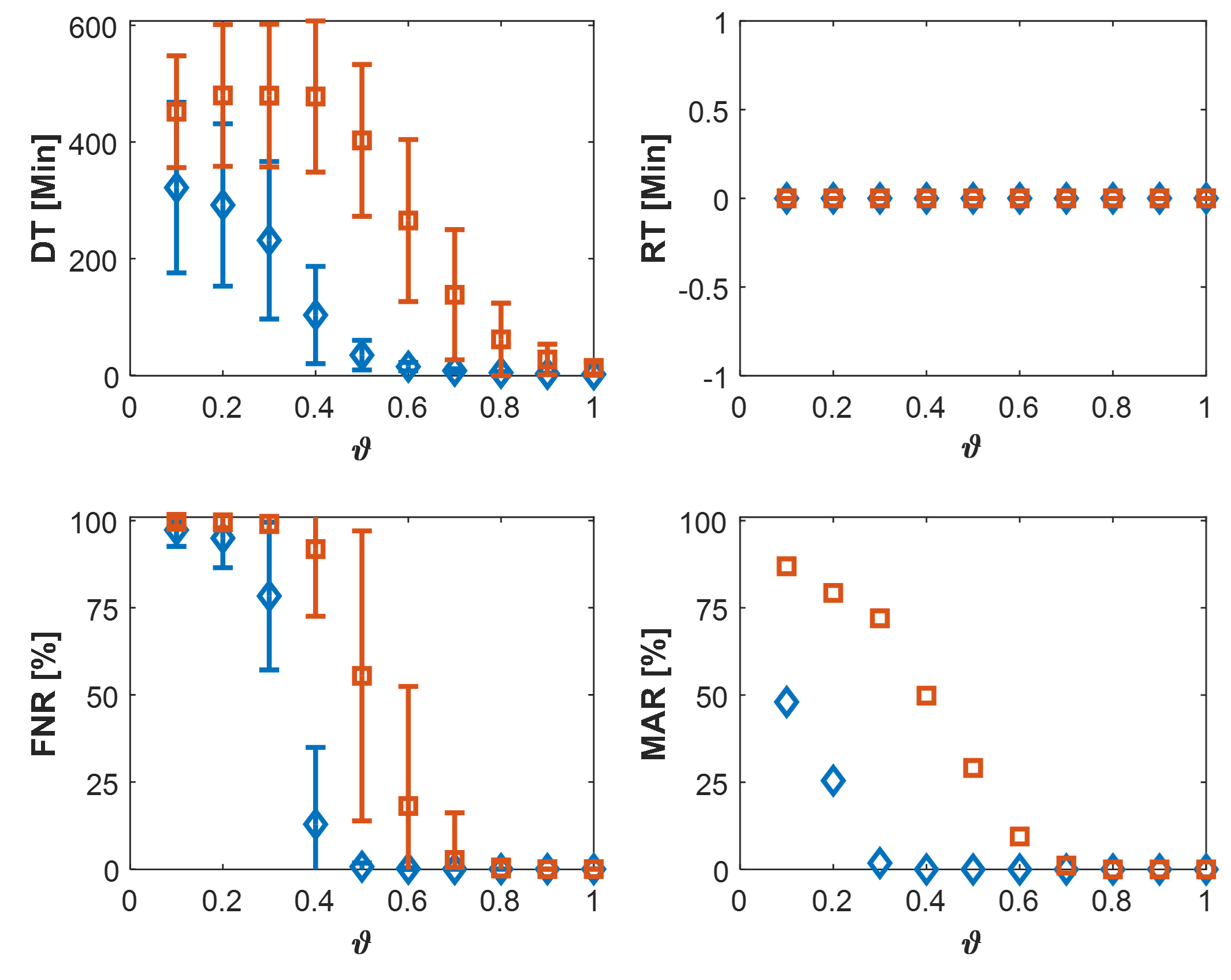}}
\caption{Variation in performance indices of direct method ($\Box$) and PCA method ($\Diamond$) with anomaly magnitude for ISC}
\label{PI_ISC}
\end{figure}

To evaluate the FPR, 24 h of nominal experimental data from twenty-five cell groups of 11 cells each were processed using both methods. The direct method and PCA method showed low average FPRs of $1.9\%$ and $2.9\%$, respectively. To explore the overall performance of the proposed anomaly detection algorithms, we tested the performance on families of synthetic anomalous data. Each anomaly was tested on all 25 cell groups, with magnitudes varying from $0.1$ to $1$ in steps of $0.1$. Figure \ref{PI_ISC} shows the DT, RT, FNR, and MAR results for the ISC. The PCA method detected ISC anomalies quicker (lower DT) and more accurately (lower FNR) than the direct method. The PCA method missed fewer anomalies overall than the direct method and detected all anomalies with $\vartheta\geq0.4$. As the anomaly spanned until the end, there was no RT in this example.

\vspace{-12pt}

\begin{figure}[H]
{\includegraphics[width=\linewidth]{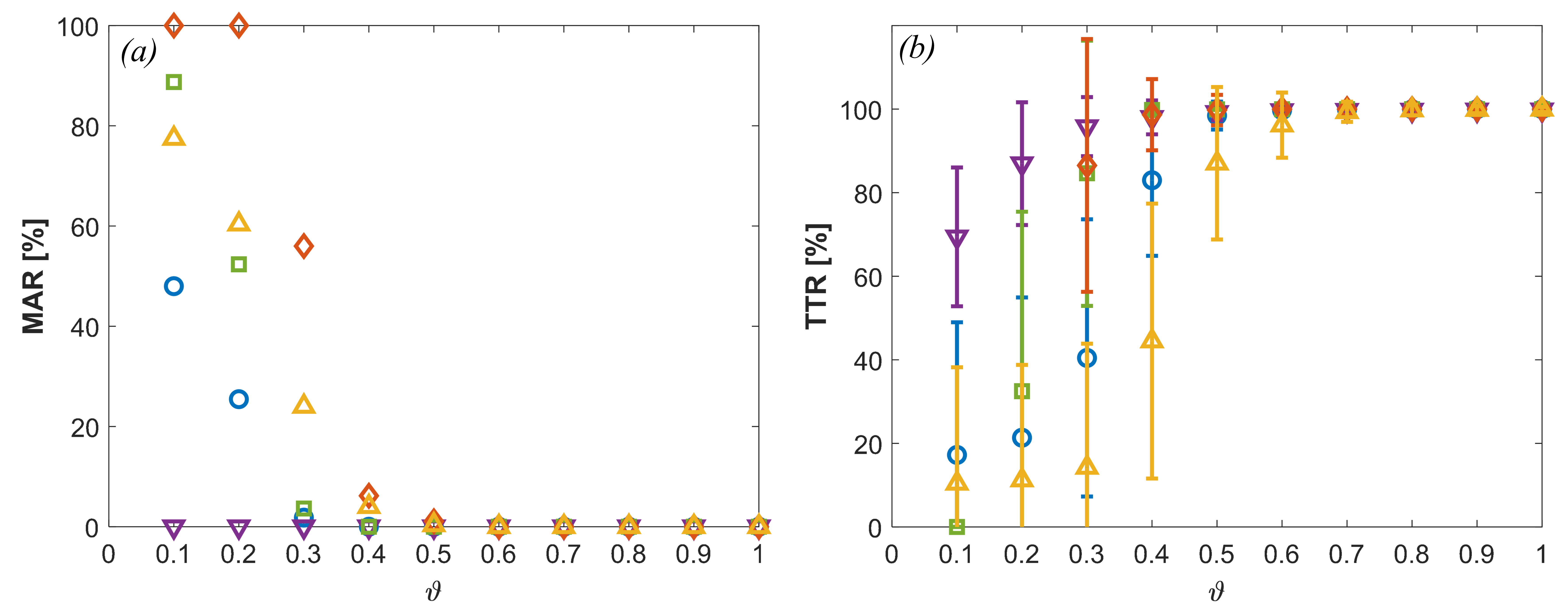}}
\caption{PCA method: (\textbf{a}) missed anomaly rate versus anomaly magnitude; (\textbf{b}) true tracing rate versus anomaly magnitude: $\bigcirc$ ISC, $\bigtriangledown$ air flow, $\Box$ loose temperature sense lead, $\Diamond$ loose voltage sense lead, and $\bigtriangleup$ voltage dropout.}
\label{MAR_TTR}
\end{figure}

Table \ref{PItable} summarizes the average DT, RT, FNR, and MAR for the five anomalies. The PCA method performed much better than the direct method in the DT, FNR, and MAR. Relative to the direct method, the PCA method improved the DT, MAR, and FNR by $56\%$, $60\%$, and $42\%$, respectively. The air$-$flow anomalies were most accurately predicted, with the lowest FNR and MAR. Loose temperature and voltage sense lead anomalies were quickly detected (low DT), as the residuals contained sudden changes. In this study, the RT is relevant only for loose voltage and temperature sense leads because the other anomalies will never recover on their own in a practical situation. Generally speaking, the RT of the PCA method was longer than that of the direct method due to the history effect in the CUSUM control chart. Figure \ref{MAR_TTR}a shows the MAR for the PCA method versus the anomaly magnitude. Voltage anomalies with deviations greater than $4$~mV and temperature anomalies with deviations greater than $0.15$ \textdegree C were detected with a zero MAR. Both methods showed similar detection rates of $\Delta V\geq26$ mV and \mbox{$\Delta T\geq2.4$ \textdegree C}, respectively. Figure \ref{MAR_TTR}b shows the TTR versus $\vartheta$ for the PCA method. The tracing accuracy increased with the increasing anomaly magnitude. Anomalous cells were correctly traced with more than $95\%$ accuracy for voltage and temperature deviations greater than $7mV$ and $0.3^oC$, respectively.

\begin{table}[H]
\centering
\caption{Average performance indices from statistical testing.}
\begin{tabular*}{\hsize}{c@{\extracolsep{\fill}} cccccccccc}
\toprule
\rule{0pt}{20pt} {\parbox{1.7cm} {\centering \textbf{Anomaly Type}}} & \multicolumn{2}{c} {\parbox{1.0cm} {\centering \textbf{ISC}}} & \multicolumn{2}{c} {\parbox{1.4cm} {\centering \textbf{Air Flow}}} & \multicolumn{2}{c} {\parbox{2cm}{\centering \textbf{Loose Temperature Sense Lead}}} & \multicolumn{2}{c} {\parbox{1.7cm} {\centering \textbf{Loose voltage Sense Lead}}} & \multicolumn{2}{c} {\parbox{1.5cm} {\centering \textbf{Voltage Dropout}}} \\[+13pt]
\midrule {\textbf{Method}}
 & \parbox{0.8cm}{\textbf{Direct}} & \parbox{0.6cm}{\centering\textbf{PCA}} & \parbox{0.8cm}{\centering\textbf{Direct}} & \parbox{0.6cm}{\centering\textbf{PCA}} & \parbox{0.8cm}{\centering\textbf{Direct}} & \parbox{0.6cm}{\centering\textbf{PCA}} & \parbox{0.8cm}{\centering\textbf{Direct}} & \parbox{0.6cm}{\centering\textbf{PCA}} & \parbox{0.8cm}{\centering\textbf{Direct}} & \parbox{0.6cm}{\centering\textbf{PCA}} \\
 \hline 
 \textbf{DT [min]} & 280 & 102 & 312 & 46 & $0.75$ & $0.72$ & 16 & 6 & 320 & 252 \\
 \textbf{FNR [\%]} & 47 & 28 & 26 & 2 & 46 & 16 & 36 & 28 & 49 & 42 \\
 \textbf{MAR [\%]} & 33 & 8 & 19 & 0 & 46 & 14 & 32 & 26 & 35 & 17 \\
 \textbf{RT [min]} & - & - & - & - & 47 & 257 & 424 & 552 & - & - \\

\bottomrule
\end{tabular*}
  \label{PItable}
\end{table}

The voltage and temperature data used in this work were collected with highly sensitive sensors with noise of less than $0.4$ mV and $0.03$ \textdegree C, respectively. Other sensing systems with lower sensitivities would not be capable of detecting $4$ mV and $0.15$ \textdegree C deviations because the thresholds would be increased to prevent false positives through the training of the CUSUM control chart on the noisy data. It is, therefore, expected that the sensitivity of voltage and temperature anomaly detection depends strongly on the quality of the sensors and data acquisition~system.

\subsection{Retraining after Balancing Events}

\begin{figure}[H]
{\includegraphics[width=0.8\linewidth]{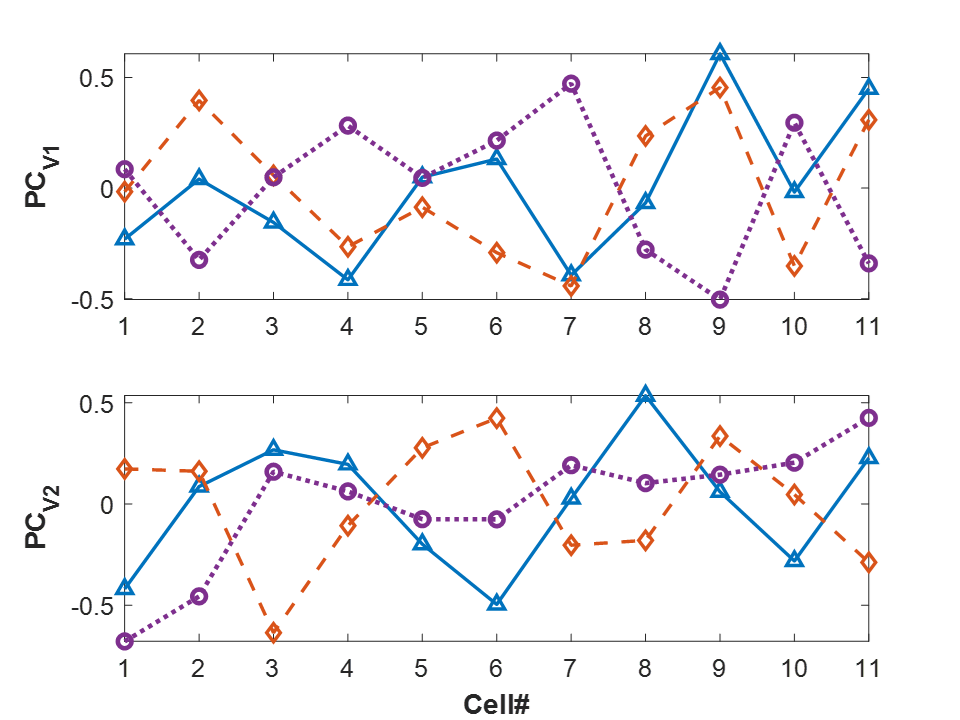}}
\caption{First two voltage principal components before balancing $\bigtriangleup$ (solid), during balancing $\bigcirc$ (dotted), and after balancing $\Diamond$ (dashed).}
\label{PCV}
\end{figure}

\begin{figure}[H]
{\includegraphics[width=0.8\linewidth]{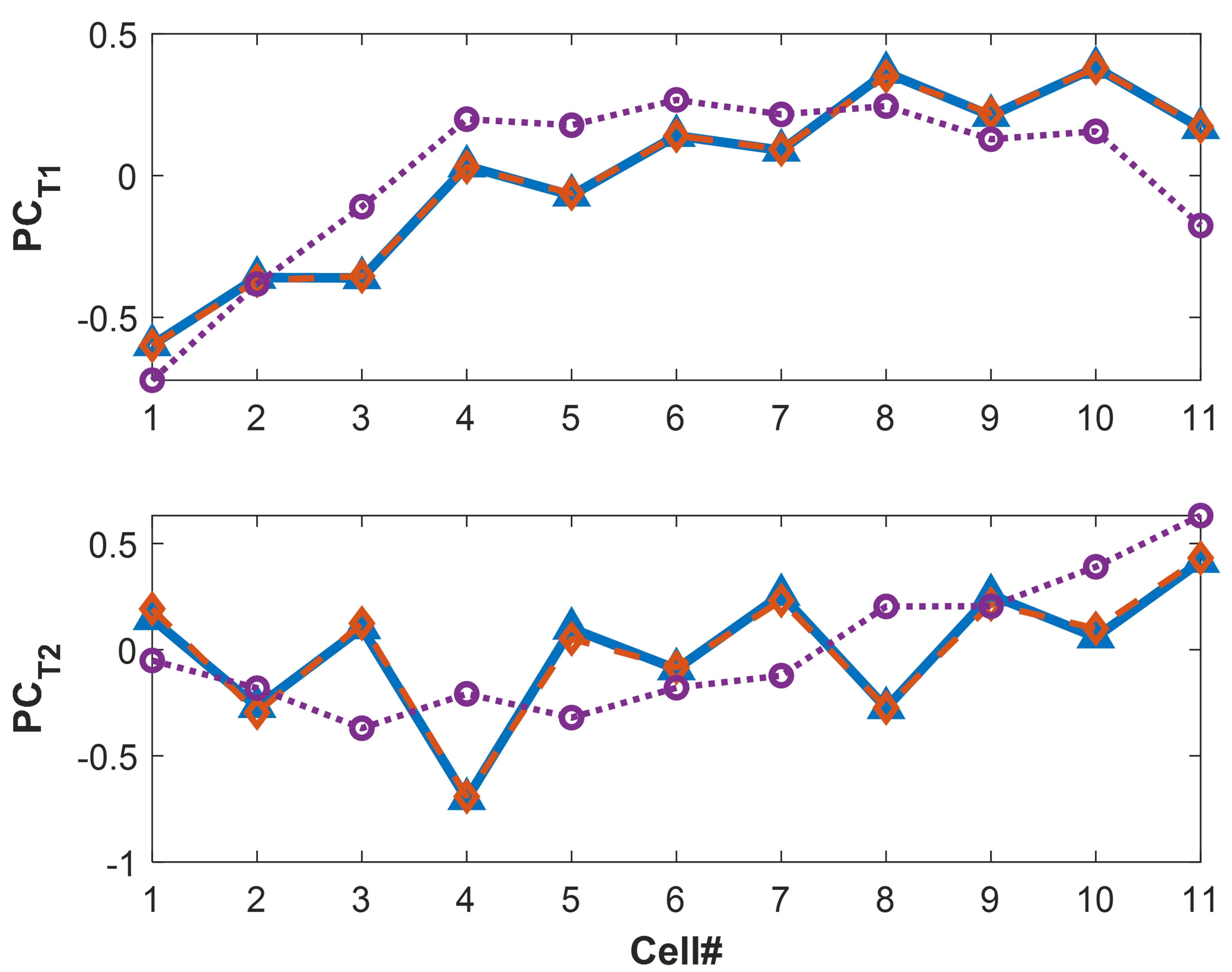}}
\caption{First two temperature principal components before balancing $\bigtriangleup$ (solid), during balancing $\bigcirc$ (dotted), and after balancing $\Diamond$ (dashed).}
\label{PCT}
\end{figure}

Balancing events occur periodically to equalize the SoC of cells in the string. Balancing is an ESC event that changes the cell-to-cell voltage relationship. The first two principal components (PCs) of the voltage and temperature before, during, and after the balancing events, are compared in Figures \ref{PCV} and \ref{PCT}, respectively. Both the voltage and temperature PCs before and during balancing did not match. Figure \ref{PCV} also shows that the cell-to-cell voltage relationships before and after balancing are different. Thus, the voltage PCA needs to be retrained after balancing to adapt to the new nominal characteristics and avoid false positives. However, Figure \ref{PCT} shows that the temperature PCs were similar before and after balancing. This was expected because balancing only changes the relative SoC of the cells, not the electrothermal cell characteristics. Thus, the temperature PCA does not need retraining after balancing events.

\vspace{-12pt}

\section{Conclusions}
This paper shows that mean-based voltage and temperature residuals for a group of similar cells can effectively detect electrical and thermal anomalies in battery systems. Mean-based residuals convert real-time voltage and temperature measurements to stationary data. These residuals are filtered and CUMSUM thresholded to detect anomalies in the direct method. Thus, the direct method detects anomalies where the temperature and/or voltage data deviate significantly from the mean. In the PCA method, PCA is used to reconstruct the normalized residuals, which are then scalarized using the RMSE as an additional step, giving the added ability to detect anomalies where the temperature and/or voltage data deviate from cell to cell. Both methods require nominal training data to establish the normalization constants, thresholds, and left singular matrix (for the PCA method). Both methods detect and trace synthetic internal short circuits, air$-$flow constrictions, and loose and broken sensor connections. False-positive rates are low ($<$3\%) and can be reduced via increased thresholds but with an increase in missed detections. Overall, the PCA method outperforms the direct method by 40--60$\%$ and is able to detect all anomalies with voltage and temperature deviations greater than $4$ mV and $0.15$ \textdegree C, respectively. Experimental ESC anomalies associated with balancing are detected within $14$ min, relying on the temperature residuals for module-level events. Voltage PCA retraining is required after cell-balancing~events.

%%%%%%%%%%%%%%%%%%%%%%%%%%%%%%%%%%%%%%%%%%
\vspace{6pt} 

%%%%%%%%%%%%%%%%%%%%%%%%%%%%%%%%%%%%%%%%%%
%% optional
%\supplementary{The following supporting information can be downloaded at:  \linksupplementary{s1}, Figure S1: title; Table S1: title; Video S1: title.}


%%%%%%%%%%%%%%%%%%%%%%%%%%%%%%%%%%%%%%%%%%
\authorcontributions{Conceptualization,
 K.B., A.J., J.B., J.P., N.B., and C.D.R.; methodology, K.B. and C.D.R.; software, K.B.; validation, K.B.; formal analysis, K.B.; investigation, J.P. and N.B.; resources, C.R.; data curation, J.P. and N.B.; writing---original draft preparation, K.B.; writing---review and editing, C.D.R.; visualization, K.B.; supervision, A.J., J.B., J.P., N.B., and C.D.R.; project administration, A.J., J.B., J.P., N.B., and C.D.R.; funding acquisition, J.B. and C.D.R. All authors have read and agreed to the published version of the manuscript.}

\funding{This research received no external funding.}

\dataavailability{Restrictions apply to the availability of these data. Data was obtained from Wabtec Corporation.}

\acknowledgments{This work was supported and funded by the Wabtec Corporation.}

\conflictsofinterest{The authors declare no conflicts of interest.} 

%%%%%%%%%%%%%%%%%%%%%%%%%%%%%%%%%%%%%%%%%%

%%%%%%%%%%%%%%%%%%%%%%%%%%%%%%%%%%%%%%%%%%
\begin{adjustwidth}{-\extralength}{0cm}

\reftitle{References}

\PublishersNote{}
\end{adjustwidth}
\end{document}